\documentclass[grl]{agujournal2019}
\usepackage{url} 
\usepackage{lineno}
\usepackage[inline]{trackchanges} 
\usepackage{soul}
\usepackage{amsmath,amssymb}
\usepackage{xr}
\draftfalse

\journalname{Geophysical Research Letters}

\begin{document}

%
%


\title{Aftershocks as a time independant phenomenon}

%
%



\authors{A. Mathey\affil{1}, J. Crassous\affil{1}, D. Marsan\affil{2}, J. Weiss\affil{3}, A. Amon\affil{1}}

\affiliation{1}{Univ Rennes, CNRS, IPR (Institut de Physique de Rennes) - UMR 6251, F-35000 Rennes, France}
\affiliation{2}{Université Savoie Mont-Blanc, CNRS, IRD, IFSTTAR, ISTerre,
Le Bourget-du-Lac, France}
\affiliation{3}{IsTerre, CNRS/Université Grenoble Alpes, Grenoble, France}




\correspondingauthor{Axelle Amon}{axelle.amon@univ-rennes.fr}





\begin{keypoints}

\item Quake sequences from laboratory experiments and seismic catalogs can be analyzed jointly in a common framework.
\item Clustering observed in laboratory and seismic sequences is not time dependent but deformation dependent.
\item Clustering properties from laboratory and seismic catalogs can be collapsed on a single master curve.

\end{keypoints}

%
%

%
%


\begin{abstract}
[ Sequences of aftershocks following Omori's empirical law are observed after most major earthquakes, as well as in laboratory-scale fault-mimicking experiments. Nevertheless, the origin of this memory effect is still unclear. In this letter, we present an analytical framework for treating labquake and earthquake catalogs on an equal footing. Using this analysis method, we show that when memory is considered to be in deformation and not in time, all data collapse onto a single master curve, showing that the timescale is entirely fixed by the inverse of the strain rate. ]
\end{abstract}

\section*{Plain Language Summary}
[ After major earthquakes, a local increase in seismic activity is generally observed. The origin of these aftershocks is still unclear, but time-dependent effects are generally invoked to explain them. We show here that time is not the relevant variable for understanding this phenomenon. By jointly analyzing catalogs of seismic faults and experiments, we show that even if these data differ by several orders of magnitude in terms of length and time scales, they can be rationalized on a single curve if we consider that the variable governing the dynamics is deformation and not time.  ]

%
%

%


%
%
%
%

\section{Introduction}
On February 6, 2023, two devastating earthquakes struck Turkey and Syria within nine hours of each other. These events are a sad reminder
of the vital importance of understanding aftershock mechanisms in order to assess the associated risk. It is well established that there
is an excess in the rate of seismicity after an earthquake, and that this excess decreases inversely with the time elapsed since the
mainshock, a law known as Omori-Utsu's Law~\cite{Omori1894,Utsu1995}. Yet the physical origin of this clustering in earthquake dynamics is still unclear.

Static stress redistribution after a mainshock seems to play a role in the increase (or decrease) of the local seismicity rate by modifying
locally the stress state of the system and thus its distance to friction thresholds~\cite{stein1994,toda1998}. Alone, the static stress redistribution is not sufficient to explain the different features observed and memory effects. In particular, several faults that experienced a significant increase in local stress still did not undergo a major earthquake, in contradiction to what has been predicted from stress transfer and average loading stress rate~\cite{scholz2019,freed2005}. Dynamic stress triggering~\cite{felzer2006decay}, as well as time-dependent effects, such as pore pressure transfers~\cite{jonsson2003}, afterslip~\cite{van2017}, or time-dependent friction laws~\cite{gross1997}, have been therefore invoked separately or combined. In particular, the state and rate friction law is appealing as it has been shown to predict the Omori's law dependence~\cite{dieterich1994}. However, the validity of transposing this law, obtained empirically on a laboratory scale, to real kilometre-scale faults with a memory effect lasting decades remains an open question~\cite{scholz1998}.

A major limitation of all those time-dependent mechanisms is that none of them can explain in an unique framework the universality of aftershock sequences. Clustering is observed after shallow earthquakes as well as after deep ruptures. The characteristics of faults vary widely in terms of the presence of water (or not) in the pores, the width and the nature of the gouge, the type of fault, the average temperature and nature of the rocks in the slip zone~\cite{kagan1991,scholz2019}. The lack of a common explanation based on a few ingredients independent of local specificity is all the more problematic that aftershock sequences following the Omori-Utsu's law have been recently observed in lab experiments on sheared, cohesion-less, ideal granular materials at low confining pressure~\cite{lherminier2019,houdoux2021}. In such systems, time-delayed redistribution of fluids or coupling with a viscous asthenosphere can not be invoked to interpret the observed clustering.



In this letter, we want to determine whether the clustering sequences observed in earthquake catalogs and in experiments are governed by the same laws and, if so, what sets the time scale of the process. To answer this question, we need to perform a direct, quantitative comparison of the clustering dynamics observed in earthquakes and in experiments, the difficulty lying in the fact that the time and length scales of these systems differ by several orders of magnitude. We propose a method of analysis which allows to study in a unique framework data coming from experiments and from seismic catalogs. Using this method of analysis, we show that it is possible to rationalize the memory effect as a deformation-dependent, and not a time-dependent, one and collapse all the data on a single master curve. This allows us to discuss the physical origin of the aftershocks as deformation-dependent processes.

In section~\ref{sec:data}, we present the two systems studied: seismic catalogs and experiments using glass beads. We detail how we estimate the mean shear rate at the level of the fault in those different systems. In section~\ref{sec:method} we present our method analysis, first the binning procedure which allows to map data from different systems in a common and objective way, second the definition of the correlation function we used to characterize quantitatively the clustering in the sequences. In section~\ref{sec:results} we apply this tool to the four systems considered and show that the time scale in all the systems is set by the inverse of the mean strain rate so that all the data can be collapsed without any further normalisation on a single power law of exponent -1. In section~\ref{sec:discussion} we discuss the results.

\section{Data}\label{sec:data}
The data used in this study come from two different systems, covering strikingly different scales: (i) earthquake catalogs from strike-slip simple (roughly linear) faults and (ii) experiments on granular materials in which slip events are measured inside a shear-band.

\subsection{Seismic catalogs}
\begin{figure}[h]%
\centering
\includegraphics[width=\textwidth]{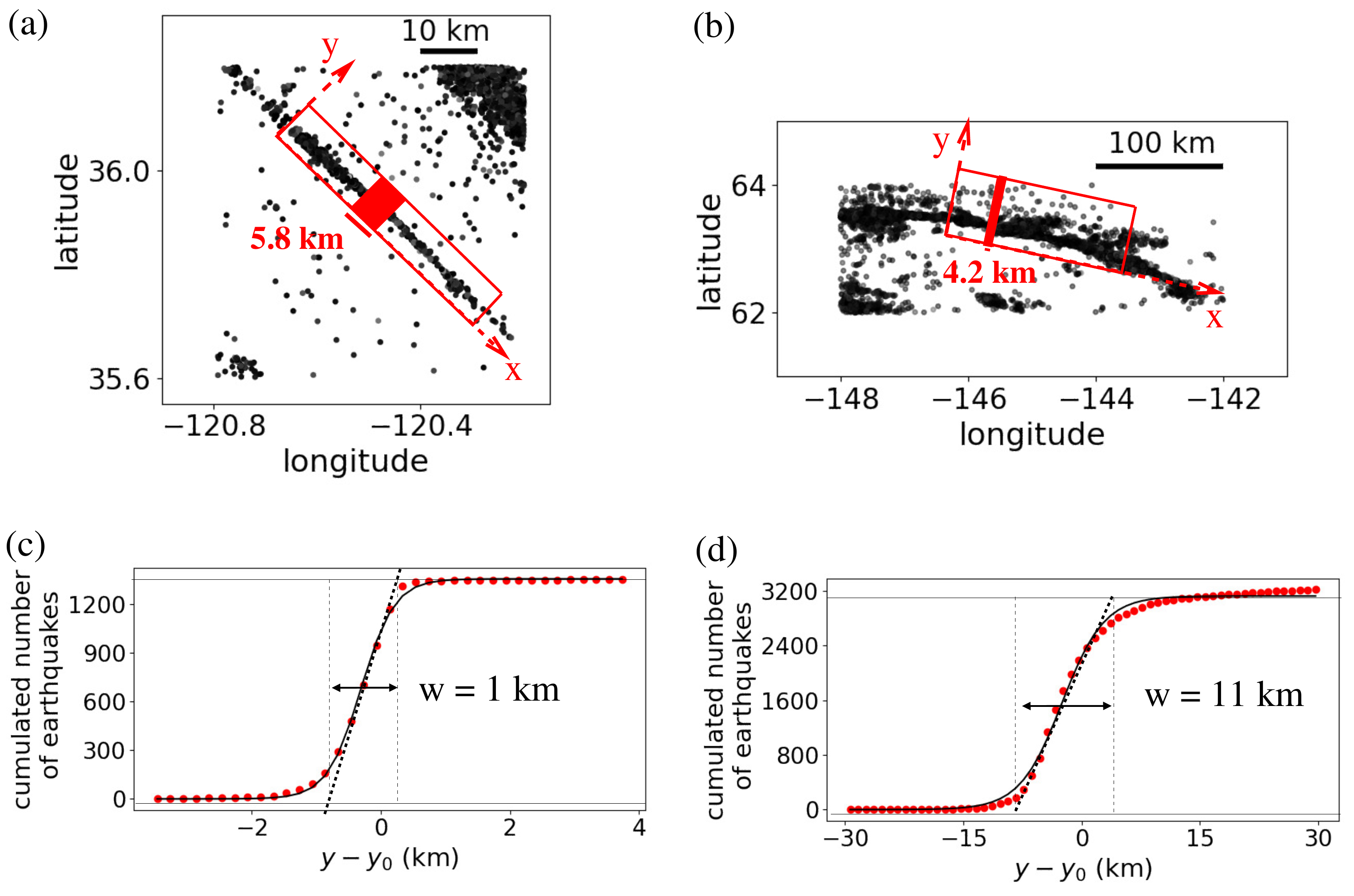}
\caption{Locations of the earthquakes used for the analysis in (a) California (Parkfield segment) and (b) Alaska (Denali fault), with local axes giving the direction of the fault and the normal to the fault. The spatial binning is shown with red boxes. Black scale: scale of the map; red scale: size $\Delta x$ of the bin. (c) (resp. (d)) Red dots: cumulative number of events in the $y$ direction integrated along the $x$-axis for map (a) (resp. (b)). Black solid line: fit using the function $f(y)=\frac{N_T}{2} \left[ 1+ \tanh\left( -2\frac{y-y_0}{w}\right)\right]$.}\label{fig:maps}
\end{figure}

We use the United States Geological Survey (USGS) earthquake catalog, targeting two almost linear strike-slip faults: the Parkfield segment of the San Andreas fault in California (latitude $\in [35.6;36.2]$, longitude $\in [-120.8;-120.2]$, from 1/1/1980 to 1/1/2024) and the Denali fault in Alaska (latitude $\in [62;64]$, longitude $\in [-148.;-142]$, from 1/1/1980 to 1/1/2024), keeping only earthquakes of magnitude larger than 2 to be sure of the completeness of the catalogs (see Fig.~\ref{fig:maps}a and b for the exact portions of the catalog studied).

The choice of those segments is motivated by the fact their simple geometry allows to estimate the mean macroscopic strain rate directly from the velocity field and the width of the fault. We use geodetic measurements of the velocity field $v$ across the fault from the literature: 24-35~mm/yr for Parkfield~\cite{platt2010real} and 5.5-15.5~mm/yr for Denali~\cite{biggs2007multi}. In both cases the uncertainty on this value is rather high. We estimate the width of the shear band by studying the cumulative number of events in the direction normal to the fault (Fig.~\ref{fig:maps}c and d). The sigmoid curves obtained can be fitted and the width $w$ of the faults estimated. We obtain $w \simeq 1$~km for Parkfield and $w \simeq 11$~km for Delani. This last value is consistent with the one that can be deduced from~\cite{biggs2007multi} which is of the order of 20~km (fig.~15 in~\cite{biggs2007multi}). Those estimates of the width of the fault vary from 10 to 50\% depending on the cut-off magnitude and the area chosen. Taking this additional uncertainty into account, we deduce the macroscopic strain rates for both faults $\dot{\varepsilon} = \frac{v}{2w}$, giving $3-6 \times 10^{-8}$~days$^{-1}$ for Parkfield and $0.4-2 \times 10^{-9}$~days$^{-1}$ for Denali.

\subsection{Experiments}

\begin{figure}[htb]%
\centering
\includegraphics[width=\textwidth]{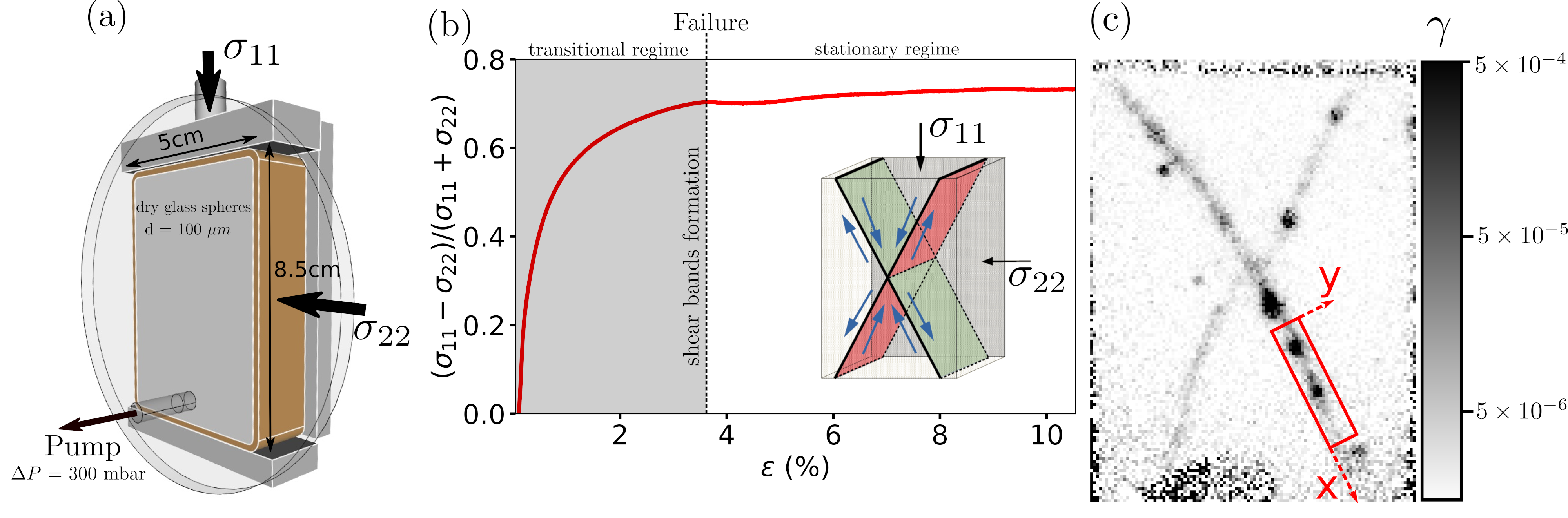}
\caption{Experiment on glass beads pack: (a) Principle of the biaxial test. Dry glass beads are enclosed in a latex membrane and submitted to a confining pressure $\sigma_{22}=30$~kPa. The sample is loaded along the '1' direction at a constant strain rate and a force sensor allows to deduce the stress $\sigma_{11}$ exerted at the top of the sample. (b) Typical loading curve representing the normalized deviatoric stress measured as a function of the overall deformation imposed. Failure occurs at about 4\% of deformation after which the stress reaches a plateau. (c) Maps of incremental deformation obtained using an interferometric method of measurement. After failure, all the deformation is localized in stationary shear bands. Deformation inside the shear bands occurs principally in the form of localized intermittent bursts. $\gamma$ is the local strain measured optically and is displayed on the map in grey levels.}\label{fig:exp}
\end{figure}

We perform biaxial tests on assemblies of non-cohesive glass micro-beads (Fig.~\ref{fig:exp}a). The sample is loaded quasi-statically at constant strain-rate. After failure, all the deformation is localized inside shear bands whose orientation is given by the Mohr-Coulomb angle. We focus on the post-failure regime, when the shear bands are stationary (Fig.~\ref{fig:exp}b). We resolve fluctuations of plasticity occurring inside the shear band using an interferometric method of deformation (see~\cite{amon2017b} for details). We have shown in previous works~\cite{houdoux2021} that those fluctuations of plasticity correspond to local slip events along our experimental fault. Those events can be sorted in a catalog using a thresholding procedure~\cite{mathey2024device}. The seismic moment associated to those events can be estimated quantitatively from the spatial size of the events and the amount of local slip. The distribution of the moments follows a Gutenberg-Richter law and time clustering of events is observed that follows the Omori-Utsu's law~\cite{houdoux2021}. The typical magnitude of the observed events are in the range -10 to -7, illustrating the huge scale gap with the earthquakes analyzed below.

In the ideal case of two symetric conjugated shear bands dividing the sample in four blocks of well-defined velocities (inset in Fig.~\ref{fig:exp}b), the strain-rate along the shear-band $\dot{\varepsilon} = \dot{\varepsilon}_{xy}$ can be deduced from the imposed compression velocity $v_1$, the angle of the shear band $\theta$ and the width $w$ of the shear band: $\dot{\varepsilon} = \frac{v_1}{4w}\sqrt{\frac{1}{\tan^2 \theta} + 1}$.

\begin{figure}[htb]%
\centering
\includegraphics[width=\textwidth]{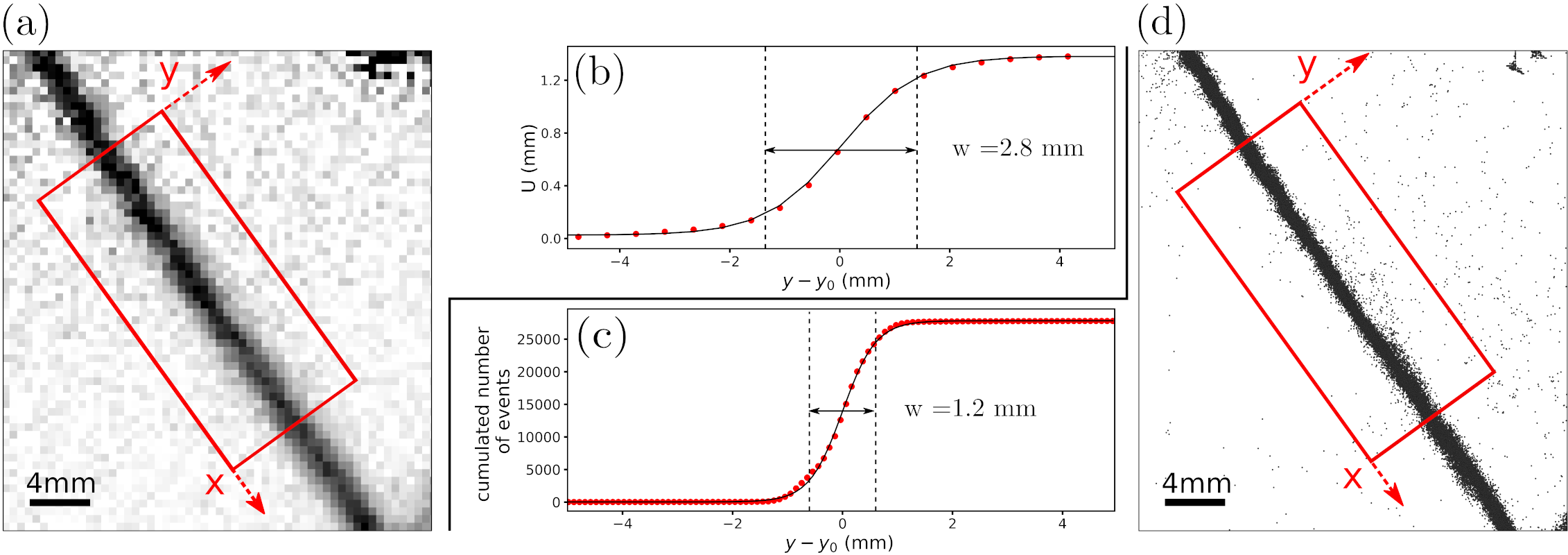}
\caption{(a) Red box: region of interest as studied in the experiment. The deformation maps have been averaged over 100 frames, showing the mean, stationary, deformation $\bar{\gamma}(x,y)$ inside the shear band. (b) (resp. (c)) Red dots:  integration of the average mean deformation $\langle \bar{\gamma} \rangle_{x}$ along the $y$ direction for map, giving the total displacement $u$ along the $x$ direction during the stationary regime. (a) (resp. cumulative number of events in the $y$ direction integrated along the $x$-axis for map (b)). Black solid line: fit using the function $f(y)=\frac{N_T}{2} \left[ 1+ \tanh\left( -2\frac{y-y_0}{w}\right)\right]$. (d) Events from the experimental catalog corresponding to the area shown in (a). }\label{fig:w_exp}
\end{figure}

We can use experiments to validate our measurement of fault width for earthquakes. Indeed, we have two ways of measuring the width of the shear band in the case of the experiment. Firstly, we can integrate the deformation $\gamma$ directly from the spatially-resolved  deformation maps (Fig.~\ref{fig:w_exp}a) to obtain the displacement $u_x$ (see Text S1 in Supporting Information and ~\cite{houdoux2021} for details). This displacement can be fitted with a sigmoid curve giving a width $w\simeq30d = 2.8$~mm for the shear band. Secondly, we can apply the same procedure as for earthquakes to the experimental catalogs (Fig.~\ref{fig:w_exp}c and d). The cumulative number of events (Fig.~\ref{fig:w_exp}c) can also be fitted by a sigmoid curve which gives a smaller width, typically half the value obtained on the raw maps. The latter method relies on the dispersion of the position of the barycenters of the events while the former method takes into account the spatial distribution of the deformation. This explain that the first method gives a larger width. Interestingly, a factor of roughly 2 is also found for the Denali's fault between our estimation of the width of the fault and the one that can be deduced from~\cite{biggs2007multi}. In conclusion, the method used to estimate the width of the shear zone from catalogs does indeed give the right order of magnitude.

We have performed experiments using glass beads of mean diameter $d\simeq90~\mu$m at two velocities of compression $v_1$: 0.25~$\mu$m/s and 2~$\mu$m/s. The Mohr-Coulomb angle and the shear band width have been measured on the spatially-resolved maps: $\theta \simeq 64^{\circ}$. The inferred values of the strain-rates are respectively $2.5 \times 10^{-5}$~s$^{-1}$ and $2.0 \times 10^{-4}$~s$^{-1}$. The uncertainty on those values is due to the simplified model used to deduce the velocity along the shear bands. In fact, the lack of symmetry in the real system and deformation within the blocks themselves result in an inhomogeneous distribution of shear strain among the different segments. We also observe during stationary regimes a slow evolution of the shape of the shear bands as well as of the mean shear strain along each segment. Those effects result in an uncertainty on the value of $\dot{\varepsilon}$ that we estimate to be of the order of 30\%.


\section{Method of analysis}\label{sec:method}
The usual data processing method of counting the excess number of events after an earthquake of a given magnitude (mainshock) treats each sequence separately, and does not allow direct quantitative comparison of data from experiments with those from field catalogs.

Our method of analysis is based on a two-step procedure. First, we map the catalogs, which consist in lists of events at irregular times $t_i$, into series defined on evenly spaced times and positions by counting the number of events taking place between $t$ and $t+\Delta t$ at a given location on a spatial grid. We call \emph{binning} this process. Second, we compute the time correlation function of the series obtained.

\subsection{Binning procedure}
We simplify the binning in space by taking advantage of the linear shape of the faults considered here. Denoting $x$ the direction along the fault and $y$ the orthogonal direction, we divide the area of interest only along the $x$ direction, in $n_x$ bins of width $\Delta x$ (fig.~\ref{fig:maps}a and b). We choose as space bin size the rupture size of the largest event of the catalog. We define an objective time bin size for the different systems using a fraction of the mean interoccurence time, defined as $\tau = \langle t_{i+1} - t_i\rangle$, where the average is taken on all the events of the catalog. As the time series have been split in $n_x$ series corresponding to different positions along the fault, the mean inter-event time at a fixed position is larger than the one of the whole catalog by a factor of the order of $n_x$. We thus fix the bin time size as $\Delta t = \frac{\tau n_x}{q}$, where $q$ is a constant that we have chosen so that $\Delta t$ corresponds to the inverse of the framerate in the experiments ($q=10$). Finally, we count the number of events $N(x,t)$ from a catalog that have occurred between the positions $x$ and $x+\Delta x$ along the fault during the time interval $[t,t+\Delta t]$. Those choices for the size of bins ensure that the mean density of events (ratio of the total number of events $N$ of the catalog to the total number of cells $n_x \times n_t$, where $n_t$ is the total duration of the catalog divided by $\Delta t$) is the same for all the systems studied.

The values of the parameters used for the different systems are given in table~S1 in Supporting Information. The size of the events are directly measured in the case of the experiments as our method provides spatially-resolved maps of deformation. In the case of the earthquakes catalogs, we use the following estimation:
$$L = 0.006 \times 10^{0.5 m}$$ 
where $L$ is  the diameter (in km) of a circular crack with 3 MPa stress drop for an earthquake of magnitude $m$~\cite{eshelby1957determination}.

\subsection{Correlation function}
To quantify the excess of activity rate compared to the mean rate, we use correlation functions of the number of events for different time lags $\delta t$. Compared to the counting method, this tool does not discriminate between mainshocks and aftershocks. It quantifies the excess of probability to observe a similar pattern after a time interval $\delta t$ for a stationary time series (see text S2 in Supporting Information).

We compute the time auto-correlation function of the function $N(x,t)$ along the fault as:
\begin{equation}
  C(\delta t) = \frac{\left< \ \overline{N(x,t) N(x,t+\delta t)} \ 
    \right>_x}{\left< \  \overline{N(x,t) N(x,t')} \ \right>_x} - 1 \label{eq:correl}
\end{equation}
where $\overline{\cdot}$ is the temporal average on the whole duration of the catalog and $\langle \cdot \rangle_x$ is the average on the position along the fault. Time series $N(x,t')$ are obtained by randomly reshuffling data
in time at a fixed position. The order of the averaging procedures (first on time, then in space) allows to take into account spatial heterogeneity along the fault (see text S3 in Supporting Information).

Note that because of the ratio which intervenes in the definition of $C$, it is not necessary to define a density of events (i.e. the number of events normalized by the width of the shear band in the $y$ position and the bin size in the $x$ direction) as such factors will cancel out. This definition of the correlation function thus allows to compare quantitatively the dynamics at different spatial scales without any further normalization.

\section{Results}\label{sec:results}

\begin{figure}[h]%
\centering
\includegraphics[width=\textwidth]{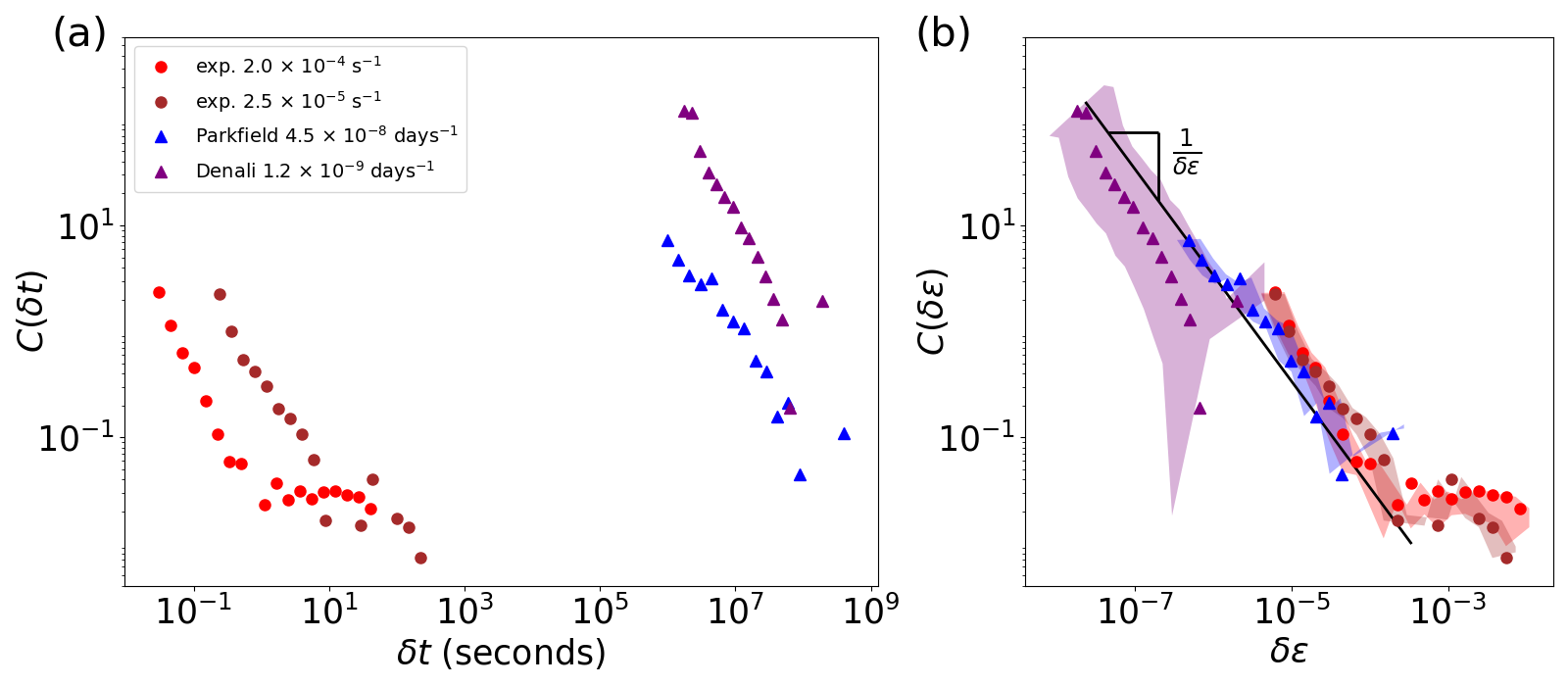}
\caption{Correlation function as defined by equation~\ref{eq:correl} for four systems: red and brown circles, experiments on sheared granular material; blue triangles Parkfield fault; purple triangles, Denali fault. (a) As a function of time lag $\delta t$, (b) as a function of deformation lag $\delta \varepsilon = \dot{\varepsilon} \delta t$ with errorbars due to the uncertainties on the estimation of $\dot{\epsilon}$ (see main text). The black line with slope -1 is shown as a visual guide. Note that the horizontal tick spacing is the same for the two plots.}\label{fig:collapse}
\end{figure}

The correlation functions $C(\delta t)$ for the four systems studied are plotted in figure~\ref{fig:collapse}a. The values have been computed for linearly spaced $\delta t$ and the resulting data have been averaged on logarithmically spaced bins (see Figure S1 in Supporting Information). The four curves follow power laws, $C(\delta t)\sim\delta t^{-\theta}$ with an exponent $\theta\simeq 1$. The time intervals over which they are defined are largely disconnected, as expected for phenomena belonging to two very different classes. The red and brown dots correspond to the two experiments done at velocities that differ by a factor of 10. The time resolution for those data is fixed by the framerate of the recording (40 and 5 fps respectively) and the typical duration of the stationary regime is of the order of one hour. The blue and purple triangles correspond respectively to the Parkfield and the Denali faults. The smallest increment $\delta t$ is fixed by the choice of the time bin size ($\approx 9.5$ and $17.4$ days respectively, see table S1 in Supporting Information).

Figure~\ref{fig:collapse}b shows the correlation functions $C$ plotted as a function of the deformation lag $\delta \varepsilon = \dot{\varepsilon} \delta t$, using the strain rate estimation computed in sec.~\ref{sec:data}. We observe that the four curves collapse approximately onto a single master curve corresponding to a power law $C(\delta \varepsilon)\sim\delta \varepsilon^{-\theta}$ with exponent $\theta\simeq 1$. Such a dependence is awaited for series obeying the Omori-Utsu law (see text S4 in Supporting Information). The $C(\delta \varepsilon)$ curves have significant uncertainties mainly linked to the difficulty to estimate the strain rates $\dot{\varepsilon}$, as discussed in sec.~\ref{sec:data}.

We have further checked the robustness of this result when the parameters of the binning procedure are changed (see Figure S2 in Supporting Information).

\section{Discussion}\label{sec:discussion}

The collapse of figure~\ref{fig:collapse}b shows that it is possible to directly and quantitatively compare the dynamics of quakes observed at time and space scales differing by several orders of magnitude by rescaling time by the strain rate. This means that the phenomena studied are truly quasi-static in the sense that time does not play an intrinsic role: the time scale in each system is fixed by the inverse of the strain rate. This shows that loss of memory in those systems is primarily governed by the amount of slip that has occurred since the reference time, not the time elapsed. Such a picture is in agreement with~\cite{stein2009} where it has been reported that the duration of aftershocks sequences varies inversely with the loading rate. In their work, Stein and Liu studied several aftershock sequences, some of which were triggered by earthquakes that occurred several hundred years ago. The sequences are located in tectonic regions governed by very different mechanisms (from intraplate to plate boundary), so that the local loading rate $v$ varies by as much as a factor of 1000 over all the sequences considered. They observe that the duration of the sequence of aftershocks $t_a$ typically varies as $1/v$.

The collapse of experiments and earthquakes on a master curve demonstrates the need to find a common origin for the clustering mechanism, which does not involved time-dependent effects. Our experiments are performed using assemblies of dry glass beads at low confining pressure. There are thus no fluid transfers, no cohesion bonds which could be damaged or healed, no crushing of beads and no thermal effects. Consequently, the available ingredients to build a model of deformation-dependent clustering are elasticity, disorder and (time-independent) friction. Such models have been extensively studied in the past in the form of assemblies of frictional blocks connected by springs~\cite{burridge1967} and in its cellular automata version~\cite{olami1992}. Those systems display an intermittent dynamics of avalanches of slip events. The size distribution of those avalanches follow a power law distribution, but, when the ingredients of the models are restricted to the previous list, they do not display time clustering~\cite{jagla2010realistic}. Interestingly, a recent theoretical work~\cite{petrelis2023} has shown that clustering can emerged in a time-independent  model if the stress field describing the internal stresses in the system has a self-affine geometry. Petrelis \emph{et al.} have shown that in the vicinity of a maximum of such a geometrical surface there is a high probability to find patches with large stress values. Said otherwise, near a site close to the threshold which will be at the origin of a mainshock, other sites also close to their thresholds exist, which could lead to aftershocks after further loading. In this picture, clustering originates from the complex structure of the internal stress field which has a correlated, self-affine geometry. The emergence of such a geometry could be the result of the combination of elastic stress redistribution and disorder.

Another point of view is based on the self-affine properties of fault surfaces themselves. It has been shown that the roughness of fault surfaces present self-affinity over several decades of length scales (from 50~$\mu$m to 50~km)~\cite{candela2012roughness}. Considering that an earthquake corresponds to a slip along such surface, the loss of memory when successive slips occur is not link to time but to the cumulative slip along the surface. This picture allows to understand why memory is governed by an amount of slip $v \times \delta t$ rather than a time lag $\delta t$. Moreover, as faults surface are self-affine, the loss of correlation of the heights between two points of the surface obeys to a power law as a function of the distance between those points~\cite{schmittbuhl1995scaling}, with an exponent which depends of the Hurst exponent of the surface. Consequently, a power law dependence can also be awaited for the excess of probability to slip at a position where a slip has occurred before, which is exactly what our correlation function $C(\delta \varepsilon)$ measures.

 
In this work we have focused on faults of very simple geometries similar to the one of our experiment in order to be able to apply exactly the same tools to the different systems and thus compare objectively and quantitatively their dynamics. A challenge is to apply the same approach to more complex faults or even to tectonic areas featuring complex hierarchies of faults. The extension of the spatial binning to a 2D grid instead of a line does not present any difficulties for larger areas. The main difficulty resides in the evaluation of the strain rate playing a role at the level of a cell of the grid which will be generally much larger than the one that can be evaluated on a large area. Such a work will be the object of another study.

\section{Conclusion}\label{sec:conclusion}
In this work, we have shown that experimental data on sheared granular media and seismic catalogs can be processed together in a unified analysis. We have shown that clustering, observed in sequences from systems of different nature and characterized by time and length scales differing by several orders of magnitude, can be rationalized when the memory effect is considered as a deformation-dependent and not a temporal effect. To our knowledge, this is the first study to analyze laboratory earthquakes jointly with real earthquakes, and to quantify clustering on a single master curve.

The central result of our paper is the fact that clustering is not a time-dependent phenomenon both in experiments and in seismic sequences. This is in essence a quasi-static phenomenon, the time scale of which is fixed by the strain rate imposed in the vicinity of the fault. This result questions interpretations of aftershock sequences relying on intrinsic time-dependent processes.

%
%
%
%


\acknowledgments
This section is optional. Include any Acknowledgments here.
The acknowledgments should list:\\
All funding sources related to this work from all authors\\
Any real or perceived financial conflicts of interests for any author\\
Other affiliations for any author that may be perceived as having a conflict of interest with respect to the results of this paper.\\
It is also the appropriate place to thank colleagues and other contributors. AGU does not normally allow dedications.


%
%



\bibliography{biblio_main}

\end{document}


%
%


\title{Supporting Information for "Aftershocks as a time independant phenomenon"}
%
%

%
%



\authors{A. Mathey\affil{1}, J. Crassous\affil{1}, D. Marsan\affil{2}, J. Weiss\affil{3}, A. Amon\affil{1}}

\affiliation{1}{Univ Rennes, CNRS, IPR (Institut de Physique de Rennes) - UMR 6251, F-35000 Rennes, France}
\affiliation{2}{Université Savoie Mont-Blanc, CNRS, IRD, IFSTTAR, ISTerre,
Le Bourget-du-Lac, France}
\affiliation{3}{IsTerre, CNRS/Université Grenoble Alpes, Grenoble, France}

%
%

%

\begin{article}

%
%

\noindent\textbf{Contents of this file}
\begin{enumerate}
\item Text S1 to S4
\item Figures S1 to S2
\item Tables S1
\end{enumerate}

\noindent\textbf{Introduction}

The supporting information provides technical details on several points. Text S1 presents the measurement of the displacement between the two sides of the shear band in the experiment. Text S2  discusses the question of the stationarity of the time series and is accompanied of Figure S1 which verifies this hypothesis for the different systems. Text S3 gives some technical details on the averaging procedure for the computation of the correlation function (eq. (1) of the main text). Text S4 gives the expected dependency awaited for the correlation function for a data series following the Omori-Utsu law. Figure S2 shows how $C$ varies when some binning parameters are varied. Table S1 gives the parameters used for the binning procedure in the main article.\\[5mm]


\noindent\textbf{Text S1. Width of the shear band.}

We can measure the displacement $u_x(y)$ between the two sides of the band using the spatially-resolved deformations maps. First we average those maps in time to obtain the mean deformation field $\bar{\gamma}(x,y)$ in a region of interest corresponding to a linear part of one band (Fig. 3a of the main article). Then, we average this value along the $x$ direction and next we integrate the deformation across the band. We thus obtain the displacement $u_x(y)$ between the two sides of the band:
$$u_x(y) = \int_{-\infty}^{y} \langle \bar{\gamma} \rangle_x \ dy.$$

\vspace{5mm}

\noindent\textbf{Text S2. Definition of the correlation function $C$}

The correlation function $C$ as defined by equation (1) is well defined in the case of a stationary data series. In that case,
$$\overline{N(x,t) N(x,t')} = \overline{N(x,t)} \; \ \overline{N(x,t')} = \overline{N(x,t)} \; \ \overline{N(x,t+\delta t)}.$$ 

\begin{figure}
\setfigurenum{S1} 
\includegraphics[width=.85\textwidth]{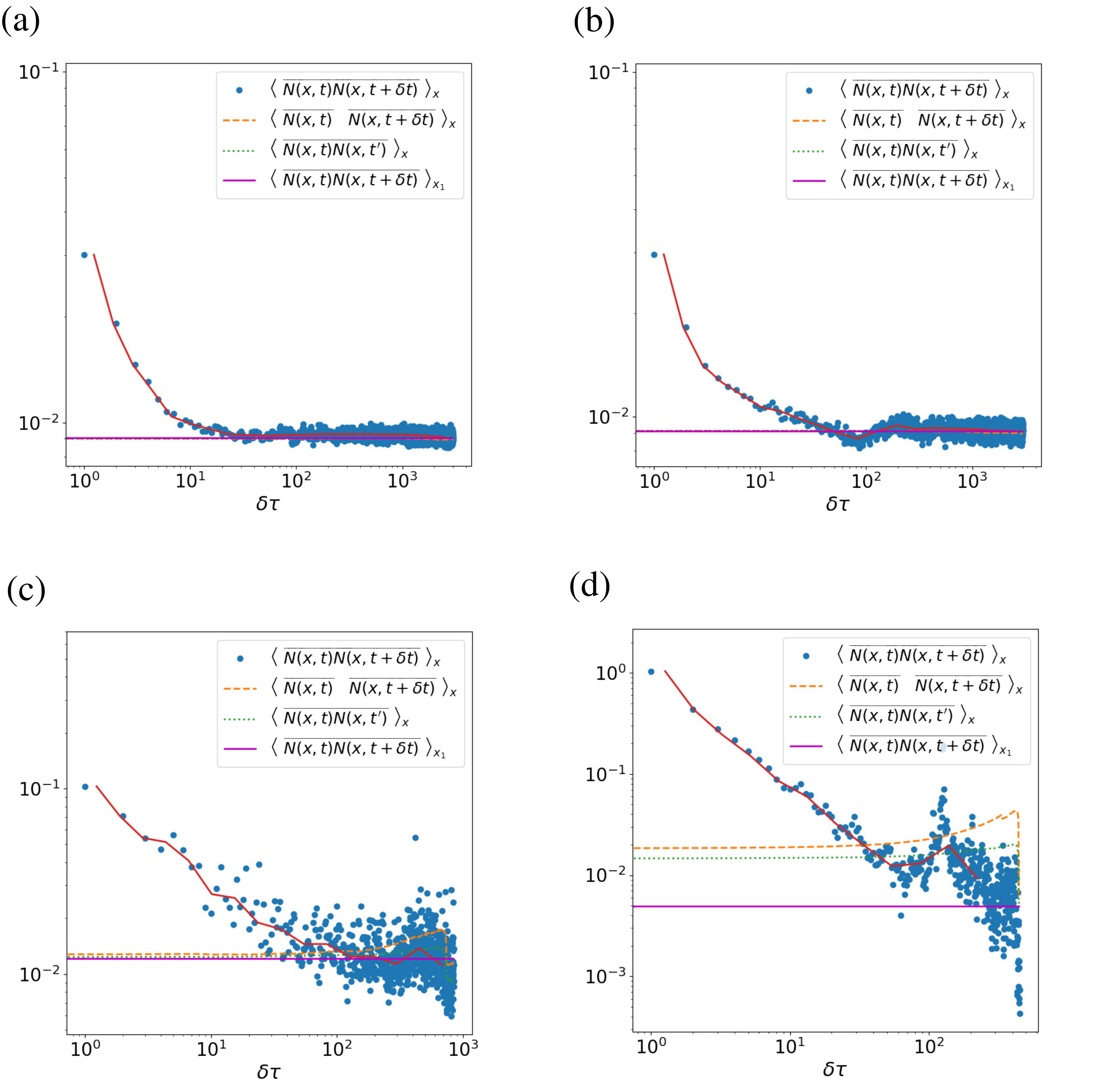}
\caption{Details of the computation of the correlation function $C$ (eq.~(1) of the article). Blue dots: $\left< \ \overline{N(x,t) N(x,t+\delta t)} \ \right>_x$; red solid line: average of the blue dots with logarithmic bins; orange dashed line: $\left< \ \overline{N(x,t)}  \ \ \ \overline{N(x,t+\delta t)} \ \right>_x$; green dotted line: $\left< \ \overline{N(x,t) N(x,t^\prime)} \ \right>_x$; magenta line: average of the ending part of the curve in blue dots (typically 1000 last points). (a) Experiment 1, (b) Experiment 2; (c) Parkfield; (d) Denali.} \label{fig:def_C}
\end{figure}

It can be verified in Figure~\ref{fig:ref_C} that it is the case for our experiments and for the Parkfield catalog. Indeed, the cross product converges towards a plateau at large lag equal to $overline{N(x,t) N(x,t')}$. In the case of the Deanli's catalog, it can be observed that the stationary regime is not reached on the duration of the catalog, probably because the aftershock sequence following the 2002 earthquake dominates the dynamics: the cross product (blue dots) are still decreasing at large values of the lag. Consequently, the values of $C$ computed from equation~(1) are only approximations of the values that would be computed if we were able to obtain a longer time series. We took into account this effect in Figure~(3)b as an additional uncertainty that we estimate by computing $C$ either with the definition of equation (1) or by using the average of values obtained for the cross product at large lag (magenta line in Figure~\ref{fig:def_C}.\\[5mm]
%


\noindent\textbf{Text S3. Averaging procedure}

The order of the averaging performed in the computation of $C$ are based on the hypothesis that heterogeneities in the spatial activity can be taken into account by expressing the number of events as a product of a function depending only on space and one depending ony on time, i.e.: 
$$N(x,t) = f(x) g(t).$$

We have then:
\begin{eqnarray*}
  \frac{\left< \ \overline{N(x,t) N(x,t+\delta t)} \ 
    \right>_x}{\left< \  \overline{N(x,t) N(x,t')} \ \right>_x}  &=& \frac{\left<f(x)^2 \ \overline{g(t) g(t+\delta t)} \ 
    \right>_x}{\left< f(x)^2 \ \overline{g(t) g(t')} \ \right>_x}\\
    &=& \frac{\left<f(x)^2\right>_x \overline{g(t) g(t+\delta t)}}{\left<f(x)^2\right>_x \overline{g(t) g(t')}}\\
    &=& \frac{\overline{g(t) g(t+\delta t)}}{\overline{g(t) g(t')}}
\end{eqnarray*}

Consequently, if the hypothesis of separation of variables hold, the order of the averages (first time at fixed space, then space) allows to take care of possible differences in the local activities.\\[5mm]

\noindent\textbf{Text S4. Correlation function of a data series following the Omori-Utsu law}

Consider a time series ot the form $n(t) = \frac{1}{t+c}$ with $c$ a constant. Then, the cross-product $X(\delta t) = \overline{n(t) \times n(t+\delta t)}$ can be computed explicitely:
\begin{eqnarray*}
  X(\delta t) &=& \frac{1}{T-\delta t} \int_0^{T-\delta t} n(t) n(t+\delta t) dt \\
          &=& \frac{1}{T-\delta t} \int_0^{T-\delta t} \frac{dt}{(t+c) (t+\delta t+c)}\\
  &=& \frac{1}{{T-\delta t}} \frac{1}{\delta t} \int_0^{T-\delta t} \left[ \frac{dt}{t+c} -
      \frac{dt}{t+\delta t +c}\right]\\
  &=& \frac{1}{{T-\delta t}} \frac{1}{\delta t} \left[ \ln \left(
      \frac{{T-\delta t}+c}{c} \right) - \ln \left( \frac{T+c}{\delta t + c} \right)
      \right]
\end{eqnarray*}

If $c \ll \delta t \ll T$,

$$X(\delta t) \simeq \frac{1}{\delta t} \left[ \frac{1}{T} \ln \left(
    \frac{\delta t}{c}\right)\right]$$
which is close to $1/\delta t$ for $\delta t \gg c$

\clearpage

\begin{figure}
\setfigurenum{S3} 
\includegraphics[width=\textwidth]{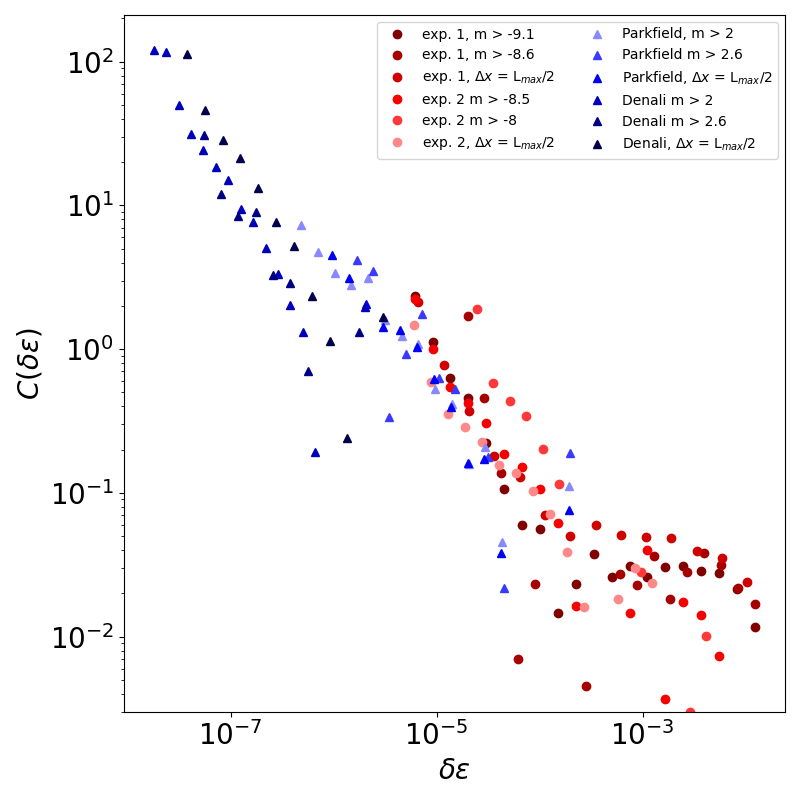}
\caption{To test the robustness of our analysis, we check the influence of the choice of parameters on the values obtained for the correlation function $C(\varepsilon)$: (i) we modified the cut-off for the minimal magnitude chosen in the catalogue, (ii) we divide $\Delta x$ by two.} \label{fig:robustesse}
\end{figure}

\clearpage

\begin{table}[h]
\settablenum{S1}
\begin{center}
\caption{Parameters chosen for the binning procedure for the different systems.}\label{tab:values}
\begin{tabular}
{c|c|c|c|c}
     system & $N$ & $\Delta x$ & $\Delta t$ & $\dot{\varepsilon}$\\
     \hline
     experiment 1 & 16696 & 5 mm (55$d$) & 0.027 s & 2.0 $\times 10^{-4}$ s$^{-1}$\\
     experiment 2 & 11217 & 5 mm (55$d$) & 0.2 s & 2.5 $\times 10^{-5}$ s$^{-1}$\\
     Parkfield & 1354 & 5.8 km & 8.2 $\times 10^{5}$ s & 5.2 $\times 10^{-13}$ s$^{-1}$\\
     Denali & 3253 & 4.2 km & 1.5 $\times 10^{6}$ s & 1.4 $\times 10^{-14}$ s$^{-1}$
\end{tabular}
\end{center}
\end{table}

\clearpage
\end{article}
\clearpage